\documentclass[3p,times,twocolumn]{elsarticle}

\usepackage{ecrc}

\volume{00}

\firstpage{1}

\journalname{Nuclear Physics B Proceedings Supplement}

\runauth{}

\jid{nuphbp}

\jnltitlelogo{Nuclear Physics B Proceedings Supplement}

\usepackage{amssymb}

\usepackage[figuresright]{rotating}

\usepackage{amsmath}
\usepackage{pstricks}
\usepackage{color} 
\usepackage{amssymb}
\usepackage{slashed}
\usepackage{graphicx,amsfonts,layout,appendix,subfigure}
\input{epsf}

\def\beq{\begin{equation}} \def\eeq{\end{equation}}
\def\beqn{\begin{eqnarray}} \def\eeqn{\end{eqnarray}}
\def\bom#1{{\mbox{\boldmath $#1$}}} \def\to{\rightarrow}
\def\nn{\nonumber}

\def\li#1{\mathrm{Li_2}\left(#1\right)}

\def\Eq#1{Eq.~(\ref{#1})}

\newcommand\KerLO{\la {\cal{P}}^{(0)}_{q_1 \gamma_2} \ra}

\newcommand\IDC{\textbf{Id}}

\newcommand\DST{D_{\rm ST}}

\newcommand\IC{\textbf{\textit{I}}}

\newcommand\CG{c_{\Gamma}}

\newcommand\g{g_{\mathrm{S}}}

\def\ep{\epsilon}

\def\wp{\widetilde P}

\def\beq{\begin{equation}}
\def\eeq{\end{equation}}
\def\beeq{\begin{eqnarray}}
\def\eeeq{\end{eqnarray}}

\def\bom#1{{\mbox{\boldmath $#1$}}}
\def\to{\rightarrow}

\newcommand{\la}{\langle}
\newcommand{\ra}{\rangle}

\def\nn{\nonumber}

\def\ID{1 \kern -.45 em 1}
\def\RS{{\scriptscriptstyle\rm R\!.S\!.}}

\def\sp{{\bom {Sp}}}
\def\ket#1{|{#1}\ra}

\def\cbet0{b_0}

\begin{document}

\begin{frontmatter}




\title{NLO QCD corrections to triple collinear splitting functions}


\author[label1,label2]{Germ\'an F. R. Sborlini}

\address[label1]{Departamento de F\'\i sica and IFIBA, FCEyN, Universidad de Buenos Aires, 
(1428) Pabell\'on 1 Ciudad Universitaria, Capital Federal, Argentina}
\address[label2]{Instituto de F\'{\i}sica Corpuscular, 
Universitat de Val\`encia - Consejo Superior de Investigaciones Cient\'{\i}ficas,
Parc Cient\'{\i}fic, E-46980 Paterna, Valencia, Spain}

\begin{abstract}
In this article, we study the triple-collinear limit of scattering amplitudes, focusing the discussion in processes which include at least one photon. To deal with infrared divergences we applied dimensional regularization (DREG) and we worked in the time-like (TL) kinematical region in order to ensure the validity of strict-collinear factorization. Both polarized and unpolarized splitting functions were obtained using independent codes, which allowed to implement a first cross-check among them. The divergent structure of all the triple-collinear splittings was compared with the Catani's formula, and we found a complete agreement. Moreover, in the polarized case, this comparison imposed additional constraints in the finite part of some master integrals (MI). The analysis of photon-started splittings led to very compact expressions, because of gauge invariance. These contributions were identified with the Abelian terms of the remaining splitting functions, which constitutes another cross-check of the results.
\end{abstract}


\end{frontmatter}


\section{Introduction}
\label{sec:Introduction}
During the last years there was an enormous progress in the computation of physical observables at higher-orders. Based on KLN theorem, we know that virtual and real contributions must be put together to obtain finite results. However, both of them contain certain divergences originated in the loop or in the phase-space integration. For this reason, it is important to properly understand the singular behaviour of scattering amplitudes.

In this article we briefly explore the collinear regime of scattering amplitudes in the context of QCD$+$QED, and describe the computation of splitting functions in the triple collinear limit. These objects control the singular behaviour of scattering amplitudes when two or more particles become collinear~\cite{Berends:1987me, Mangano:1990by}. When working in the time-like (TL) kinematical region, strict-collinear factorization~\cite{Collins:1989gx} guarantees the universality of splitting functions and their independence of the non-collinear particles \cite{Catani:2011st,Forshaw:2012bi}.

For the double collinear limit, splitting functions were first introduced in Ref. \cite{Altarelli:1977zs}. They have been computed at one-loop~\cite{Bern:1998sc, Bern:1999ry, Bern:1993qk, Bern:1994zx,Bern:1995ix,Kosower:1999rx,Sborlini:2013jba} 
and two-loop level~\cite{Bern:2004cz, Badger:2004uk, Vogt:2005dw, Vogt:2004mw, Moch:2004pa, Moch:2014sna}, both for amplitudes and squared matrix-elements. The multiple collinear limit has been studied since it is an essential ingredient of ${\rm N}^k$LO hadronic computations. Many tree-level multiple collinear splittings were computed \cite{Campbell:1997hg, Catani:1998nv, DelDuca:1999ha, Birthwright:2005ak, Birthwright:2005vi, Catani:1999ss}, although higher-orders corrections are not fully known. At one-loop level, there were only some partial results for $q \to q \bar{Q} Q$ \cite{Catani:2003vu}. In Refs. \cite{Sborlini:2014mpa, Sborlini:2014kla}, we gave a full description of triple-collinear splitting functions at one-loop level for processes which involve at least one photon.

The outline of this paper is the following. In Section \ref{sec:properties} we establish the notation and describe Catani's formula. Besides that, we briefly describe the computational techniques applied to obtain the results. In Section \ref{sec:photons} we discuss the photon-started triple-collinear splitting functions, focusing in the polarized case. We make some comments on the structure of the results, specially about gauge invariance properties. After that, we describe the remaining splitting functions in Section \ref{sec:QCDstarted} and present the conclusions in Section \ref{sec:conclusion}.

\section{Kinematics and other properties of the collinear limit}
\label{sec:properties}
In the most general configuration, let's consider an $n$-particle process where $m$ particles become collinear at the same time. Momenta are labelled as $p_i$ and $C=\{1,2, \ldots, m\}$ denotes the set of collinear particles. Partons are considered massless, so $p_i^2=0$. Subenergies are defined as $s_{i j} = 2 \, p_i \cdot p_j$ and $s_{i,j} = \left(p_i +p_{i+1} + \ldots + p_{j} \right)^2 = p_{i,j}^2$. Strict-collinear factorization is fulfilled in the TL region, i.e. $s_{ij} \geq 0$ for every $i,j \in C$, so we performed all the computations in this kinematical regime. Also, it is suitable to use a Sudakov-like parametrization to describe the collinear momenta. For this purpose we introduce the light-like vectors $\wp^{\mu}$ and $n^{\mu}$, so that
\beq
\wp^{\mu} = p_{1,m}^{\mu} - \frac{s_{1,m}}{2 \; n \cdot \wp} \, n^{\mu} \, ,
\label{ptilde}
\eeq
corresponds to the collinear direction and $n^\mu$ describes how the collinear limit is approached, with $n\cdot\wp = n\cdot p_{1,m}$. Also, we define
\beq
z_i = \frac{n \cdotp p_i}{n\cdot \wp}~, \qquad i \in C~,
\eeq
as the longitudinal momentum fractions, which fulfil $\sum_{i\in C} z_i = 1$.

Besides kinematics, factorization properties become manifest when working in the \textit{light-cone gauge} (LCG) \cite{Pritchard:1978ts, Curci:1980uw}. In spite of some technical difficulties\footnote{Computations in the LCG involve dealing with spurious divergences, which are originated by the presence of linear denominators inside Feynman integrals.}, the ghosts decouple from the theory and the gluons only have physical polarizations. Since the collinear limit involves the presence of almost on-shell virtual states, LCG guarantees that internal lines can be expressed in terms of physical particles. Thus, keeping the most singular contribution when $s_{1,m} \to 0$, we obtain
\beqn
\nn \ket{{\cal A}\left(p_1, \ldots, p_n\right)} &\simeq& \sp_{a \to a_1 \ldots a_m}(p_1 ,\ldots, p_m; \wp) 
\\ &\times& \ket{{\cal A}(\wp, p_{m+1}, \ldots, p_n)}~,  
\label{FACTORIZACIONLOmultiple1}
\eeqn
where $\sp_{a \to a_1 \ldots a_m}$ is the splitting amplitude. When the parent parton is a vector particle $V$, it is possible to remove its polarization vector $\epsilon_{\mu}(\wp)$ and define amputated splitting amplitudes. The polarized splitting functions are obtained from the tensor product of two amputated splitting matrices, i.e.
\beqn
\nn \textbf{P}^{\mu \nu}_{V \to a_1 \ldots a_m} &\equiv& \left( \frac{s_{1,m}}{2 \;\mu^{2\ep}} \right)^{m-1} \; \left(\sp^{\mu}_{V \to a_1 \ldots a_m}\right)^{\dagger} 
\\ &\times& \sp^{\nu}_{V \to a_1 \ldots a_m} + \, {\rm h.c. }\, ,
\label{PpolarizedNLOdefinition1}
\eeqn
where there is an implicit sum over the colors and spins of the external partons, and we perform an average over the parent parton's colors. $\textbf{P}^{\mu \nu}_{V \to a_1 \ldots a_m}$ keeps all the spin information of the parent parton, thus it allows to have a complete description of the collinear limit. To obtain the unpolarized splitting, we just contract with $d_{\mu \nu}$ and divide by the number of polarizations,
\beqn
\la \hat{P}_{V \to a_1 \cdots a_m} \ra &=& \frac{1}{\omega} \, d_{\mu \nu}(\wp,n) \,  \textbf{P}^{\mu \nu}_{V \to a_1 \ldots a_m} \, \, ,
\label{EcuacionDESPOLARIZA} 
\eeqn
where $\omega=2(1-\epsilon)$ and
\beqn
d_{\mu \nu}(\wp,n) &=& -\eta_{\mu \nu}^{\DST} + \frac{\wp_\mu n_{\nu}+n_\mu \wp_{\nu}}{n \cdot \wp} \, ,
\eeqn
is the LCG gluon propagator in the context of DREG \cite{Bollini:1972ui,'tHooft:1972fi}. Notice that we use $\eta_{\mu \nu}^{\DST}$ instead of the $4$-dimensional metric because we work in the CDR scheme \cite{'tHooft:1972fi}.

In an analogous way, we can extend \Eq{PpolarizedNLOdefinition1} for fermion-started processes. Instead of removing a polarization vector, we amputate the spinor $u(\wp)$ and obtain an object with open fermion chains. In the $4$-dimensional theory, helicity conservation implies
\beqn
{\textbf{P}}_{q \to a_1 \ldots a_m}(s,s') &=& \omega_q \, \delta_{s,s'} \, \la \hat{P}_{q \to a_1 \ldots a_m} \ra  \, ,
\label{FermionPOLvsUNP}
\eeqn
so we only require to compute the unpolarized splittings started by quarks. However, there is a subtlety related with the concept of \textit{helicity conservation} in DREG. Considering $\DST=4-2\epsilon$, there are contributions originated by helicity-violating interactions, so non-diagonal terms should be added to \Eq{FermionPOLvsUNP}. However, after removing IR/UV poles, non-diagonal terms in spin space are proportional to $\epsilon$. Since we compute the ${\cal O}(\epsilon^0)$ corrections, these additional terms are neglected here\footnote{A further discussion about this issue is available in Ref. \cite{Sborlini:2013jba}, using the double-collinear limit to make explicit computations.}.

Working in the TL-region \cite{Catani:2011st, Forshaw:2012bi}, the polarized-splitting functions are rank-$2$ tensors that depend only on the collinear momenta (i.e. $p_i^{\mu}$ with $i \in C$), the quantization vector $n^{\mu}$ and the $\DST$-dimensional metric. So, it is necessary to write all the possible tensorial structures and define a basis. For instance,
\beqn
f_{1}^{\mu\nu} &=& \eta_{\DST}^{\mu \nu} \, ,
\\ f_{1+i}^{\mu\nu} &=& \tilde{p}_{\sigma_1(i),\sigma_2(i)}^{\mu \nu} \ \ i \in \left\{1,\ldots,\Delta_1\right\} \, ,
\\ f_{1+j+\Delta_1}^{\mu\nu} &=& \tilde{p}_{j,m+1}^{\mu \nu} \ \ \ \ \ \ j \in \left\{1,\ldots,m\right\} \, ,
\\ f_{2+\Delta_1+m}^{\mu\nu} &=& \tilde{p}_{m+1,m+1}^{\mu \nu} \, ,
\\ f_{2+\Delta_1+m+i}^{\mu\nu} &=& \bar{p}_{\rho_1(i),\rho_2(i)}^{\mu \nu} \ \ i \in \left\{1,\ldots,\Delta_2\right\} \, ,
\\ f_{2+\Delta_1+\Delta_2+m+i}^{\mu\nu} &=& \bar{p}_{j,m+1}^{\mu \nu} \ \ \ \ \ \ j \in \left\{1,\ldots,m\right\} \, ,
\eeqn
with
\beqn
\tilde{p}_{i,j}^{\mu\nu} &=& p_i^{\mu}p_j^{\nu}+p_j^{\mu}p_i^{\nu} \, ,
\\ \bar{p}_{i,j}^{\mu\nu} &=& p_i^{\mu}p_j^{\nu}-p_j^{\mu}p_i^{\nu} \, ,
\\ \Delta_1 &=& m(m+1)/2 \, ,
\\ \Delta_2 &=& m(m-1)/2 \, ,
\eeqn
where we define $p_{m+1}^{\mu}=n^{\mu}$ to simplify the notation. Here $\sigma$ is a rearrangement of $m$ elements including the possibility of repeating them, while $\rho$ is just a permutation of $m$ elements. Notice that for a generic $m$-collinear process there are $\upsilon=((m+1)^2+1)$ elements in the tensorial basis. Besides that, the first $2+m+\Delta_1$ elements of the proposed basis are symmetric under the exchange $\mu \leftrightarrow \nu$, and the remaining are antisymmetric.

Since we are interested in the triple-collinear limit, we can adapt the previous basis for the case $m=3$. So, we have $\upsilon=17$ and the splitting function is expanded as
\beqn
\textbf{P}^{\mu \nu}_{V \to a_1 a_2 a_3} &=& \sum_{j=1}^{\upsilon} \, A_j f_j^{\mu \nu} \, .
\label{Descomposition2}
\eeqn
In the following, we will use the same notation applied in Ref. \cite{Sborlini:2014kla}. To obtain the coefficients $A_j$, we introduced the kinematic matrix $M$,
\beq
(M)_{ij} = f_i^{\mu \nu} {(f_j)}_{\mu\nu} \equiv (M_{\rm sym} \otimes M_{\rm asym})_{ij} \, ,
\eeq
that is written as the direct product of a $11 \times 11$ symmetric matrix ($M_{\rm sym}$) times a $6 \times 6$ antisymmetric one ($M_{\rm asym}$). It is worth noticing that $\det (M_{\rm sym}) = -8 \epsilon \Omega^5$, so $M_{\rm sym}$ becomes singular in the limit $\epsilon \to 0$. This is expected since $4$ vectors in a $3$-dimensional vector space are always linearly dependent.

Due to the fact that splitting functions are contracted with physical polarization vectors, only a subset of the tensorial structures $f_j^{\mu\nu}$ give non-trivial contributions. In other words, since $\epsilon(\wp,n)\cdot \wp =0=\epsilon(\wp,n) \cdot n$, we can cancel $n^{\mu}$ and replace $p_3^{\mu}=-p_1^{\mu}-p_2^{\mu}$ every time that $\mu$ corresponds to a polarization vector's index. In consequence, using Cramer's rule, only the relevant coefficients are extracted, which increases the computational performance. So, we obtain \cite{Sborlini:2014kla}
\beqn
\textbf{P}^{\mu \nu}_{V \to a_1 a_2 a_3} &=& \sum_{j=1}^{4} \, A^{\rm sym}_j f_j^{\mu \nu} \, + \, A^{\rm asym} f_{12}^{\mu \nu} \, ,
\label{ExpansionPmunuTRIPLE1}
\eeqn
with $f_1^{\mu \nu}=\eta^{\mu \nu}_{\DST}$, $f_2^{\mu\nu}=s_{1,3}^{-1}\tilde{p}_{1,1}^{\mu\nu}$, $f_3^{\mu\nu}=s_{1,3}^{-1}\tilde{p}_{1,2}^{\mu\nu}$, $f_4^{\mu\nu}=s_{1,3}^{-1}\tilde{p}_{2,2}^{\mu\nu}$ and $f_{12}^{\mu\nu}=s_{1,3}^{-1}\bar{p}_{1,2}^{\mu\nu}$.

\subsection{Divergent structure in DREG}
As it is well-known, loop-corrections to splitting functions exhibit IR/UV divergences. According to Catani's formula at one-loop \cite{Catani:2003vu}, splitting functions are expanded as
\beqn
\nn \sp^{(1)}_{V \to a_1 \dots a_m} &=& \IC^{(1)}_{V \to a_1 \dots a_m} \, \sp^{(0)}_{V \to a_1 \dots a_m} 
\\ &+& \sp^{(1)\,{\rm fin.}}_{V \to a_1 \dots a_m}~,
\label{DescomposicionSP1}
\eeqn
where
\beeq 
\nn &&\IC^{(1)}_{V \to a_1 \dots a_m} = \CG \, \g^2 \, \left( \frac{-s_{1, m} -i0}{\mu^2}\right)^{-\ep} \, 
\\ \nn && \Bigg\{ \frac{1}{\ep^2}
\sum_{i,j=1 (i \neq j)}^{\bar{m}} \;{\bom T}_i \cdot {\bom T}_j
\left( \frac{-s_{ij} -i0}{-s_{1, m} -i0}\right)^{-\ep} \nn \\
&+&
\frac{1}{\ep^2}
\sum_{i,j=1}^{\bar{m}} \;{\bom T}_i \cdot {\bom T}_j
\;\left( 2 - \left( z_i \right)^{-\ep} -\left( z_j \right)^{-\ep} \right) \nn \\
\nn &-& \frac{1}{\ep} 
\left( \sum_{i=1}^{\bar{m}} \left( \gamma_i - \ep {\tilde \gamma}_i^{\RS} \right)
- \left( \gamma_V - \ep {\tilde \gamma}_V^{\RS} 
\right) \right.
\\ &-& \left. \frac{\tilde{m}-2}{2} \left( \beta_0 - \ep {\tilde \beta}_0^{\RS}
\right) \right) \Bigg\}~,
\label{sp1divAA}
\eeeq
and $\sp^{(1)\,{\rm fin.}}_{V \to a_1 \dots a_m}$ does not contain $\epsilon$-poles. Here $\CG$ is the $d$-dimensional one-loop volume factor, ${\bom T}_i$ is the color charge operator associated with parton $i \in C$ and $\bar{m}$ ($\tilde{m}$) counts the number of collinear QCD partons (total QCD particles in the process, including the parent parton). In particular, we appreciate that $\tilde{m}=\bar{m}$ in collinear splittings which are started by non-QCD partons. Also, it is possible to express the insertion operator in terms of a $c$-number, i.e. $\IC^{(1)}_{V \to a_1 \dots a_m} = I^{(1)}_{V \to a_1 \dots a_m} \IDC$ because color algebra is closed when $\tilde{m} \leq 3$.

Besides that, all the scheme dependence is also controlled by \Eq{sp1divAA}, up to ${\cal O}(\epsilon^0)$. In fact, it is regulated by the coefficients ${\tilde \beta}_0^{\RS}$ and ${\tilde \gamma}_i^{\RS}$, which are zero for CDR scheme. This property is very useful to simplify our expressions because we are interested in ${\cal O}(\epsilon^0)$ corrections to the NLO contributions to the splitting functions. So, working with $\sp^{(1)\,{\rm fin.}}_{V \to a_1 \dots a_m}$ up to ${\cal O}(\epsilon^0)$, we get rid of any scheme dependence in the final result. 

\subsection{Organization of the results}
Due to the fact that the divergent structure is predicted by \Eq{sp1divAA}, we can subtract it and work with the finite remainder. For this reason, using \Eq{PpolarizedNLOdefinition1} we obtain 
\beqn
\nn \textbf{P}^{(1)\,{\rm fin.},\mu \nu}_{V \to a_1 \ldots a_m} &=& \left( \frac{s_{1,m}}{2 \;\mu^{2\ep}} \right)^{m-1} \left(\sp_{V \to a_1 \ldots a_m}^{(0),\mu}\right)^{\dagger} 
\\ &\times& \sp_{V \to a_1 \ldots a_m}^{(1)\,{\rm fin.},\nu} \, .
\label{EquacionDescomposicion2a}
\eeqn
In the triple-collinear limit, starting from \Eq{ExpansionPmunuTRIPLE1} and using \Eq{EquacionDescomposicion2a}, we obtain
\beqn
\textbf{P}^{(1)\,{\rm fin.},\mu \nu}_{V \to a_1 a_2 a_3} &=& c \left[ \sum_{j=1}^{4} A^{(1)\,{\rm fin.}}_j f_{j}^{\mu \nu} \, + A^{(1)\,{\rm fin.}}_5 f_{12}^{\mu \nu} \right] \,
\label{EquacionDescomposicion2}
\eeqn
where $c=c^{a \to a_1 \cdots a_m}$ is a process-dependent normalization factor. Because all the processes analysed in this article involve a quark-antiquark pair in the final state, they can be written in the generic form $V \to q_1 \bar{q}_2 V_3$ where $V$ and $V_3$ are vector-like particles. Thus, the corresponding polarized splitting functions are symmetric both under the exchange $\mu \leftrightarrow \nu$ and $1 \leftrightarrow 2$. In consequence, we can exploit this symmetry considerations to simplify the final results and impose consistency-checks in intermediate steps of the computation.

Using the traditional Feynman diagram approach, we write the contributing amplitudes, classify them according to the irreducible denominators involved and apply integration-by-parts (IBP) identities \cite{Chetyrkin:1981qh, Laporta:2001dd}. We end up with a list of master integrals (MI) multiplied by rational coefficients. So, we expand the result up to ${\cal O}(\epsilon^0)$ and verify that $\textbf{P}^{(1)\,{\rm fin.},\mu \nu}$ does not contain $\epsilon$-poles (which constitutes another consistency-check). Finally, we classify the finite terms according to their transcendental weight: in consequence, we obtain
\beqn
A^{(1)\,{\rm fin.}}_j &=& \sum_{i=0}^2 {\cal C}_j^{(i)}  \, + (1 \leftrightarrow 2) \,    \ {\rm for} \ j\in\left\{1,3\right\}  , \,
\\ A^{(1)\,{\rm fin.}}_2 &=& \sum_{i=0}^2 {\cal C}_2^{(i)}  \, ,
\\ A^{(1)\,{\rm fin.}}_5 &=& \sum_{i=0}^2 {\cal C}_5^{(i)}  \, - (1 \leftrightarrow 2)  \, ,
\eeqn
where ${\cal C}_j^{(i)}$ includes only functions of transcendental weight $i$.

\section{Photon splittings}
\label{sec:photons}
In order to better understand the structure of triple collinear splitting functions and its NLO corrections, let's start considering photon-initiated processes. There are two non-trivial configurations at tree-level: $\gamma \to q \bar{q} \gamma$ ($\bar{m}=2$) and $\gamma \to q \bar{q} g$ ($\bar{m}=3$). Introducing the function
\beqn
\nn {\cal P}^{\mu\nu}&=& \frac{1}{x_1 x_2} \left(\eta^{\mu\nu} \left(\epsilon x_1 (1-x_3)-(1-x_1)^2\right) \right.
\\ \nn &+& \left. 2(\epsilon-1) \, f_2^{\mu \nu}+2\epsilon \, f_3^{\mu \nu}\right)
\\ &+& (1 \leftrightarrow 2) \, ,
\eeqn
the LO polarized splitting functions for these processes are given by
\beqn
\textbf{P}^{(0),\mu \nu}_{\gamma \to q_1 \bar{q}_2 \gamma_3} &=& e_q^4 g_e^4 C_A \,  {\cal P}^{\mu\nu} \, ,
\label{POLKERNELLOA-qqbAbis}
\\ \textbf{P}^{(0),\mu \nu}_{\gamma \to q_1 \bar{q}_2 g_3} &=& e_q^2 g_e^2 \g^2 C_A C_F \,  {\cal P}^{\mu\nu} \, .
\label{POLKERNELLOA-qqbgbis}
\eeqn
As expected, the color structure is very simple and they share the same kinematical dependence. Here we used the notation
\beq
x_i = (-s_{jk} -i0)/(-s_{1,3} -i0) \, ,
\eeq
with $(i,j,k)$ a permutation of $\{1,2,3\}$.

When considering NLO corrections, we can easily appreciate that $\textbf{P}^{\mu \nu}_{\gamma \to q_1 \bar{q}_2 \gamma_3}$ contains the Abelian part of $\textbf{P}^{\mu \nu}_{\gamma \to q_1 \bar{q}_2 g_3}$. In fact, we found
\beqn
\left. \frac{\textbf{P}^{\mu \nu}_{\gamma \to q_1 \bar{q}_2 g_3}}{\textbf{P}^{\mu \nu}_{\gamma \to q_1 \bar{q}_2 \gamma_3}} \right|_{C_A\to0} &=& \frac{\g^2 C_F}{g_e^2 e_q^2} \, ,
\eeqn
which constitutes a cross-check between these splittings. If we define $D_A=2C_F- C_A$, this relation allows us to write
\beqn
{\cal C}_{j}^{(i,\gamma \to q \bar{q} g)} &=& C_A {\cal C}_{j}^{(i,C_A)} + D_A {\cal C}_{j}^{(i,D_A)} \, ,
\eeqn
with ${\cal C}^{(i,D_A)}={\cal C}^{(i,\gamma \to q \bar{q} \gamma)} /2$.

On the other hand, contributions of weight $0$ and $1$ are a bit lengthy so we will not show explicit expressions for them in this paper. However, it is worth noticing that they are independent of $z_i$. Weight $2$ contributions are also independent of $z_i$, and they are given by
\beqn
\nn {\cal C}_{1}^{(2)} &=& -2 {{\cal F}_1} \left(\frac{1-x_1}{x_2^2}+\frac{2 x_1-5}{x_2}+\frac{x_2-2}{x_1} \right.
\\ &+& \left. \frac{2}{x_1 x_2}+2\right) \, ,
\\ \nn {\cal C}_{2}^{(2)} &=& -\frac{2 {{\cal F}_2} \left(2 x_1^2+2 x_1 (x_2-1)+(1-x_2)^2\right)}{x_1^3 x_2}
\\ &-&  \frac{2 {{\cal F}_1} \left(2 x_2^2+(1-x_2)^2\right)}{x_1 x_2^3}\, ,
\\ {\cal C}_{3}^{(2)} &=& -\frac{4  {{\cal F}_1} \left((1-x_2)^2-x_1\right)}{x_1 x_2^3}\, ,
\eeqn
\beqn
{\cal C}_{5}^{(2)} &=& -\frac{4  {{\cal F}_1} x_3}{x_1 x_2^2} \, ,
\eeqn
for $\gamma \to q \bar{q} \gamma$ and
\beqn
{\cal C}_{1}^{(2,C_A)} &=& - \frac{{{\cal F}_3} (1-x_1)^2}{x_1 x_2} \, ,
\\ {\cal C}_{2}^{(2,C_A)} &=& -\frac{2 {{\cal F}_3}}{x_1 x_2}\, ,
\eeqn
for the $C_A$ part of $\gamma \to q \bar{q} g$. In the previous expressions, we used ${\cal F}_i={\cal R}\left(x_i,x_3\right)$ with $i=\{ 1,2 \}$ and
\beqn
\nn {\cal R}\left(x_i,x_j\right) &=& \zeta_2-\li{1-x_i}-\li{1-x_j}
\\ &-& \log{x_i}\log{x_j} \, .
\label{DefinicionFUNCIONR}
\eeqn
This function is related with the finite part of the \textit{standard} scalar box integral (i.e. \textit{without LCG denominators}). Its presence is related with the fact that photon-initiated splitting functions at one-loop can be expressed using only standard bubbles and boxes. Because the parent parton is a color singlet, we can attach the splitting amplitude to a colorless fermionic line and build a gauge invariant scattering amplitude. So, the one-loop computation can be performed using a covariant gauge, which avoids the presence of Feynman integrals with LCG propagators. Or, in other terms, if we compute the NLO correction to the splitting function in LCG, then LCG integrals vanish.

It is interesting to appreciate that unpolarized photon-started splitting functions depend on $z_i$, although the polarized ones are completely independent of $n^{\mu}$. We conclude that the whole $z_i$-dependence is introduced through the contraction with $d_{\mu \nu}(\wp,n)$ while performing the average over the polarizations of the parent parton.

\section{QCD started splittings and further checks}
\label{sec:QCDstarted}
There are three processes started by QCD partons which include at least one photon and have non-vanishing tree-level contributions. In one hand, we have $q \to q \gamma \gamma$ and $q \to q g \gamma$, which were computed in Ref. \cite{Sborlini:2014mpa} for the unpolarized case. On the other hand, $g \to q \bar{q} \gamma$ was computed in Ref. \cite{Sborlini:2014kla} for the polarized case. Tree-level quark-started splittings are given by
\beeq
\nn \la \hat{P}^{(0)}_{q_1 \gamma_2 \gamma_3} \ra &=& \frac{e^4_q g^4_e}{x_2}\left(\frac{\KerLO}{z_2}\left(1+ \frac{(1+x_2) z_1}{x_3}\right) \right.
\\ \nn &+& \left. \vphantom{\frac{z_1 (2 (x_1+1) x_3-(1-x_1))}{2 (1-x_1) x_3}} (\Delta -1) \left(\Delta  \left(x_1-\frac{z_1 (x_1+1)}{1-x_1}+\frac{z_1}{2 x_3}\right) \right.\right.
\\ \nn &+& \left.\left. \vphantom{\frac{z_1 (2 (x_1+1) x_3-(1-x_1))}{2 (1-x_1) x_3}} x_3+z_1-z_2+2 \right) \right)  
\\ &+& \, (2 \leftrightarrow 3)\, ,
\eeqn
\beqn
\la \hat{P}_{q_1 g_2 \gamma_3}^{(0)} \ra &=& C_F \, \frac{\g^2}{e^2_q g^2_e}  
\la \hat{P}_{q_1 \gamma_2 \gamma_3}^{(0)} \ra \, ,
\eeqn
where $\Delta=\epsilon$ in CDR and $\Delta=0$ in FDH/HV. On the other hand, working with CDR,
\beqn
\textbf{P}^{(0),\mu \nu}_{q_1 \bar{q}_2 \gamma_3} &=& \frac{e_q^2 g_e^2 \g^2}{2} \,  {\cal P}^{\mu\nu}  \,  ,
\label{POLKERNELLOg-qqbAbis}
\eeqn
is the gluon-started polarized splitting function.

The explicit NLO corrections are shown in Refs. \cite{Sborlini:2014mpa, Sborlini:2014kla}, and are not included here because of their size. It is worth making some comments about the structure of those results. In first place, we can appreciate that $\gamma \to q \bar{q} \gamma$ and $q \to q \gamma \gamma$ are diagrammatically related under the exchange $P \leftrightarrow 2$, although it is not possible to establish a crossing-like transformation to link them: parent parton is off-shell while outgoing collinear particles are on-shell. On the other hand, quark-started processes lead to expressions that are more complicated than vector-started ones. This is associated with the constraints imposed by the projection over the on-shell physical polarization vector $\epsilon_{\mu}(\wp)$. 

Besides that, many consistency checks have been applied to our computations. First of all, polarized and unpolarized splittings were calculated using two independent codes and we were able to recover the unpolarized case after contracting with $d_{\mu \nu}(\wp,n)$. In second place, for every process, we compared the divergent structure with the one predicted by Catani's formula: we found a complete agreement in all the configurations. Besides that, we explored the Abelian limit of those splittings which contains gluons. This consists in extracting a global normalization factor and taking the limit $C_A \to 0$, $N_f\to 0$. Explicitly, we found
\beqn
\left(q \to q g \gamma \right)_{C_A \to 0, N_f\to 0} &\approx& \left(q \to q \gamma \gamma \right) \, ,
\\ \left(\gamma \to q \bar{q} g \right)_{C_A \to 0, N_f\to 0} &\approx& \left(\gamma \to q \bar{q} \gamma \right) \, ,
\\ \left(g \to q \bar{q} \gamma \right)_{C_A \to 0, N_f\to 0} &\approx& \left(\gamma \to q \bar{q} \gamma \right) \, ,
\eeqn
for both polarized and unpolarized splittings. This is the expected behavior based on a naive Feynman diagram analysis.

Finally, we would like to point out that Catani's formula also imposes constraints on the finite part of some MIs. In particular, for the $g \to q \bar{q} \gamma$ polarized splitting, we expanded the LCG-box integral
\beqn
\nn && I^{\rm box}_{\rm LCG} = \int_q  \, \frac{1}{q^2 \, t_{2q}\, t_{23q}\, t_{123q} \, n \cdot q} =  \CG \g^2
\\ &\times& \left( \frac{-s_{1,3} -i0}{\mu^2}\right)^{-\ep} \left(\frac{B_0}{\epsilon^2} + \frac{B_1}{\epsilon} + B_2 \right) \, ,
\eeqn
and forced the cancellation of the single $\epsilon$-poles, after subtracting those predicted by \Eq{sp1divAA}. We obtained
\beqn
\left(B_2 + S_{1 \leftrightarrow 2}\left(B_2\right) \right) \, + 2 \, {\cal D}(x_i,z_i)  &=& 0 \, ,
\label{EcuacionBOX}
\eeqn
where ${\cal D}(x_i,z_i)$ is a function which only involves rational combinations of weight $2$ functions. Moreover, ${\cal D}(x_i,z_i)$ is expressed as a combination of bubbles, triangles and \textit{standard} boxes. So, this procedure allows to compute the symmetric ${\cal O}(\epsilon^0)$-terms of LCG-box integrals using simpler ones. Also \Eq{EcuacionBOX} imposes a cross-check among the MIs used along our computations and the expressions obtained for the splitting functions.

\section{Conclusions and outlook}
\label{sec:conclusion}
Splitting functions control the singular behavior of scattering amplitudes in the collinear limit. In the TL-region, they are process-independent and depend on the collinear particle momenta and quantum numbers only. 

In this paper, we briefly describe the computation of NLO QCD corrections to the triple-collinear splitting functions, for processes that involve at least one photon. We calculated both polarized and unpolarized splittings using independent codes. This was important to implement consistency-check among both sets of expressions: using \Eq{EcuacionDESPOLARIZA} we recover unpolarized splittings from the polarized ones. Besides that, we compare the divergent IR/UV structure with the expected behavior according to Catani's formula. We found a complete agreement and this allowed us to subtract all the $\epsilon$-poles from the results.

The study of photon-initiated processes led us to very compact expressions. The simplifications are caused by gauge-invariance, since the off-shell particle that undergoes the collinear splitting does not carry any color charge. QCD-started splittings are more complicated but it is still possible to use photon-started ones to both simplify and cross-check the results.

In a forthcoming article, we will present all the remaining triple-collinear splitting functions in QCD, up to NLO in the strong coupling. The knowledge of these objects is crucial to obtain full NNNLO hadronic cross-sections, that is the next accuracy frontier.

  \subsection*{Acknowledgments}
I would like to thank Daniel de Florian and Germ\'an Rodrigo for their extremely valuable contribution to this project.
This work is partially supported by UBACYT, CONICET, ANPCyT, the
Research Executive Agency (REA) of the European Union under
the Grant Agreement number PITN-GA-2010-264564 (LHCPhenoNet),
by the Spanish Government and EU ERDF funds
 (grants FPA2011-23778 and CSD2007-00042
Consolider Ingenio CPAN) and by GV (PROMETEUII/2013/007).












\end{document}